\documentclass[fleqn,usenatbib]{mnras}
\usepackage{array}
\usepackage{anyfontsize}

\newcounter{RowNum}
\newcommand{\rownumber}{\stepcounter{RowNum}\arabic{RowNum}.}
\newcolumntype{N}{>{\rownumber}c}

\usepackage{newtxtext,newtxmath}
\usepackage[T1]{fontenc}
\usepackage{ae,aecompl}
\usepackage{supertabular,booktabs}
\usepackage{hyperref}

\usepackage{graphicx}	

\newcommand{\numax}{\mbox{$\nu_{\rm max}$}}

\newcommand{\tess}{{\em TESS\/}}
\newcommand{\dsct}{\texorpdfstring{\mbox{$\delta$~Scuti}}{delta Scuti}}
\newcommand{\dscts}{\texorpdfstring{\mbox{$\delta$~Scuti stars}}{delta Scuti stars}}
\newcommand{\gdor}{\mbox{$\gamma$~Dor}}
\newcommand{\cd}{\mbox{d$^{-1}$}}

\newcommand{\bprp}{\mbox{$G_\text{BP}-G_\text{RP}$}}

\newcommand{\fmin}{\mbox{$f_\text{F}$}}

\usepackage{longtable}
\usepackage{booktabs} 
\graphicspath{{./}{figures/}}
\newif\ifarxiv
\arxivfalse 

\title[ZAMS II]{Characterizing bright $\delta$ Scuti pulsators using TESS light curves: II.~Pulsation amplitude and energy distributions}

\author[Mani et al.]{Prasad Mani$^1$,
Timothy R. Bedding$^1$,
Simon J. Murphy$^2$,
Daniel Hey$^3$, and
Tom Love$^2$
\\
$^1$Sydney Institute for Astronomy, School of Physics, University of Sydney, NSW 2006, Sydney, Australia \\
$^2$Centre for Astrophysics, University of Southern Queensland, Toowoomba, QLD 4350, Australia\\
$^3$ Institute for Astronomy, University of Hawaii, Honolulu, USA
}

\date{}

\pubyear{2026}

\begin{document}
\label{firstpage}
\pagerange{\pageref{firstpage}--\pageref{lastpage}}
\maketitle

\begin{abstract}
Pulsations in $\delta$ Scuti stars exhibit a wide range of excited frequencies and amplitudes, but the underlying principles governing their amplitude distribution remain unclear. We revisit this problem using a global, energy-based perspective. For a complete sample of 3200 bright $\delta$ Scuti stars observed by TESS, integrated power spectral density (PSD) of the Fourier spectrum allows us to measure the overall pulsation signal. From this, we derive a proxy for the total pulsation energy. We uncover a bimodality among higher-frequency delta Scuti stars in their integrated PSD, possibly suggesting the presence of two distinct pulsation regimes. In addition, we identify an apparent upper limit to the total pulsation energy, indicating a constrained energy budget for these pulsations. This feature is not evident when stars are characterized solely by their dominant mode amplitude. 
\end{abstract}

\section{Introduction}
The \dscts\ are predominantly p-mode pulsators \cite[][]{Bowman+Kurtz2018, Antoci++2019, Kurtz2022} located at the intersection of the main sequence with the classical instability strip (IS). They are A-F type, usually fast rotating \citep[see][]{Wang2025} intermediate mass stars whose pulsations are driven by the heat engine mechanism \citep[][]{stellingwerf1978,Guzik2021}. Also known as the $\kappa$ mechanism, this operates by blocking radiation with increased opacity in the partial He~II ionization zone during the star's compression phase. Many properties of \dsct\ pulsations are now becoming increasingly well understood, including the location and boundaries of the instability strip (IS) \citep[][]{Dupret2004, Dupret2005, Murphy2019, Mani2025}. Differences in pulsation behaviour depending on metallicity or peculiarity have also been explored, such as in SX Phe \citep[][]{nemecetal2017,Antoci++2019} and $\lambda$ Boo stars \citep{paunzenetal2002a,Murphy2020-lambda-boo} in the metal-poor case, and Am stars in the metal-rich case \citep{durfeldt-pedrosetal2024}. Recently there has been a resurgence in interest in mode amplitudes, thanks to homogeneous all-sky photometry from the Transiting Exoplanet Survey Satellite (TESS; \citealt{Ricker2015}), allowing characterisation of High Amplitude \dscts\ \citep[HADS;][]{Jia2024, Niu2024, Yang2025, Rodon2026} and the low-amplitude pulsation content alike \citep{Barcelo-Forteza2024}. Similarly, studies of hybrid stars, exhibiting both \dsct\ (p-mode) and $\gamma$ Doradus (g-mode) pulsations are gaining pace \citep[][]{Grigahcene2010, Sanchez2017, Skarka++2022, Skarka2024, Kliapets2025}.

Recent advances in modelling \dscts\ \citep[][]{Murphy2023, Scutt2023, Gautam2026} have not yet translated into a full understanding of their pulsations, largely because of the rich and complex distribution of modes in their Fourier amplitude spectra. To revisit, these stars exhibit pulsations spanning a wide range of frequencies ($\simeq$5--80\ \cd) and amplitudes ($\simeq$10--10,000\ \text{ppm}), complicating a detailed interpretation. Recent progress has been made thanks to the discovery of \dscts\ displaying regular patterns in the Fourier spectra \citep[see][]{Michel2017, suarez++2014, Bedding2020, Singh2025}, enabling seismic inferences of stellar properties. 

Our previous paper \citep[][hereafter Paper~I]{Mani2025} studied 444 bright \dscts\ ($G<7$) that lay within one magnitude of the zero-age main sequence (ZAMS). Here, we expand the sample and focus on pulsation amplitudes.
Research on the amplitudes of \dscts\ has mainly focussed on the dominant pulsation mode \citep[e.g.,][]{suarez2002, Balona2018, Bowman+Kurtz2018, Murphy2019, Bedding2023, Read++2024}, largely neglecting the contribution of the full mode spectrum. 
Here, we analyze pulsations by studying their integrated power spectral density (PSD) and total mode energy. Measuring mode amplitudes by integrating the PSD is routinely used in the study of solar-like oscillations \citep[e.g.][]{Kjeldsen2005, Kjeldsen2008, Huber2011}.

\section{Data analysis}

\subsection{Sample selection}\label{section:data analysis}

A combination of Gaia astrometry \citep[Gaia DR3,][]{Gaia-De-Ridder++2023} and TESS \citep[][]{Ricker2015} photometry provides an excellent opportunity for ensemble study of \dscts. The following cuts in various Gaia DR3 \citep[][]{Gaia-De-Ridder++2023} parameter space were made to obtain our sample.
\begin{enumerate}
    \item Gaia apparent magnitude $G<8$: favours sample completeness as detection of \dsct\ pulsations in fainter stars becomes harder due to increased photon noise \citep{Read++2024, Mani2025, Berry2025}.
    \item Gaia colour $0.05 \le \bprp\ \le 0.60$ and absolute magnitude $M_G > -1$: restricts stars close to the classical instability strip where \dscts\ are mostly found.
    \item Fractional parallax error $\displaystyle \sigma_\pi/ \pi < 0.1$: avoids stars with inaccurate measurements of distance, and hence absolute brightness.
    \item Gaia \texttt{fidelity\_v2} $> 0.75$: avoids stars with spurious parallaxes \citep[see][]{Rybizki++2022}. 
\end{enumerate}
We used ADQL query in the Gaia archive~\footnote{\url{https://gea.esac.esa.int/archive/}} with the above filters to obtain a catalogue of stars, and then downloaded their TESS light curves. Since \dscts\ are often poorly sampled with 1800-s data, we only considered 120-s, 200-s, or 600-s cadence light curves, which have high enough Nyquist frequencies for a proper spectral analysis. We only used SPOC \citep[Science Processing Operating Center,][]{Caldwell2020} light curves, prioriziting shorter cadence where available. We obtained 6810 stars with SPOC 120-s , 919 stars with TESS-SPOC 200-s, and 1887 stars with TESS-SPOC 600-s. The amplitude spectra for these stars were computed using Lomb-Scargle algorithm, with an oversampling factor of 10.

\begin{figure}
\includegraphics[width=\linewidth]{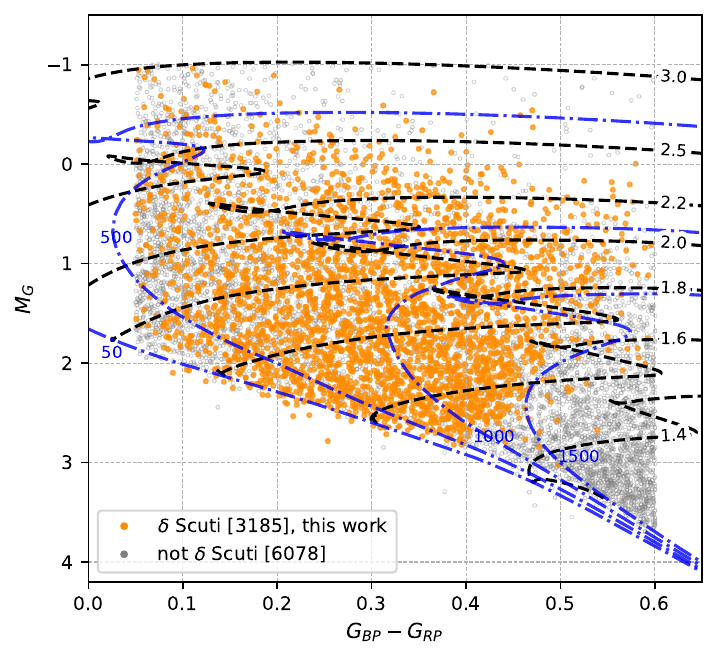}
\caption{Colour-magnitude diagram of our sample, showing \dscts\ in orange. The curves show MIST \citep{Choi2016} evolutionary tracks (blacked dashed lines, labelled in solar mass units) and isochrones (blue dot-dashed lines, labelled in Gyr), both for a rotation rate of $v/v_\text{crit} = 0.4$.}
\label{fig:CMD_plain}
\end{figure}

Eclipsing and contact binary stars were identified and discarded from our sample by either cross-matching with the \citet{Prsa2022} and Paper~I catalogues, or by checking their amplitude spectra by eye. This left us with 9263 stars in total. These stars span a distance range of 15--630 pc. We labelled a star ``\dsct'' if its amplitude spectrum had at least one strong peak [signal-to-noise (S/N) > 5] in the p-mode regime. The latter was defined with respect to the expected fundamental-mode frequency, \fmin, obtained from the empirical relation in \citet{Barac++2022}, further discussed in Section~\ref{section:PL}. To define the p-mode regime, we used a threshold of 0.66\fmin, because the literature reveals that many \dscts\ pulsate slightly below their \fmin\ \citep{Ziaali2019, Jayasinghe2020,  Barac++2022, Mani2025}. We defined S/N to be the ratio of the highest peak and the mean of white noise, where white noise was computed in 95--100 \cd\ for 120-s cadence and 67--72 \cd\  for 200- and 600-s cadences. We made a conservative choice of S/N > 5 as our threshold to strike a balance between sample purity (avoiding constant stars with random high noise peaks) on one hand, and possibly missing out on low-amplitude \dscts\ on the other hand.
The final sample of \dscts\ that we use in this work is shown in the colour-magnitude diagram (CMD) in Fig.~\ref{fig:CMD_plain}.

\subsection{Amplitudes of the dominant peaks}\label{section: A_1 and f_1}

\begin{figure*}
\includegraphics[width=\linewidth]{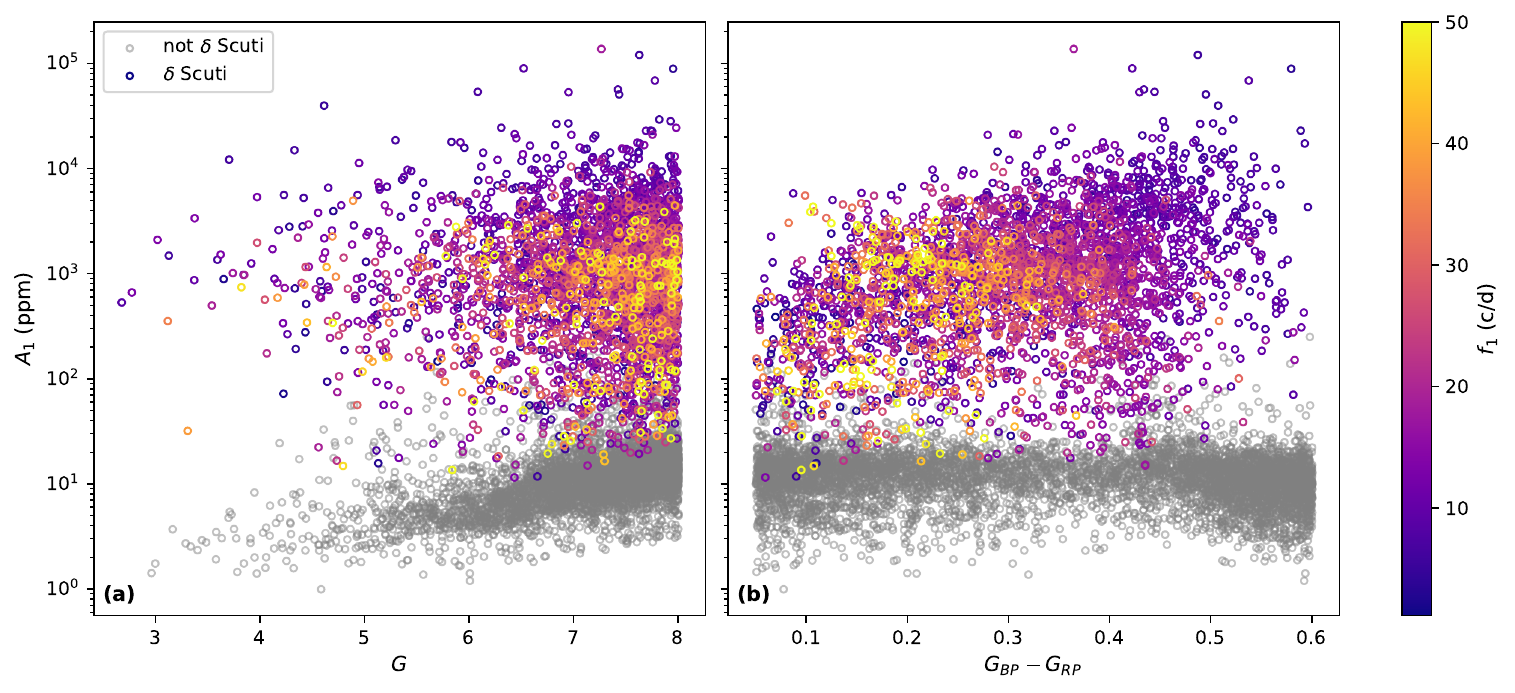}
\caption{(a): Maximum amplitude $A_1$ vs Gaia apparent magnitude $G$, and (b) vs \bprp. Stars are colour-coded by $f_1$. As very few stars have high $f_1$ values we saturate the colourbar at 50 c/d.}
\label{fig:amp_vs_bprp}
\end{figure*}

Following previous works, we first measured the amplitude ($A_1$) and frequency ($f_1$) of the dominant mode in each \dsct\ star. For ``not \dsct'', we measured the highest noise peak above 30 \cd\ to avoid mistakenly identifying any low frequency variability as a noise peak.
Fig~\ref{fig:amp_vs_bprp}~(a) shows $A_1$ vs $G$. The rise in amplitude for ``not \dsct'' (grey circles) towards fainter apparent magnitude indicates the increasing white noise floor. 
We see from the slight overlap of \dsct\ and ``not \dsct'' at lower-right that we are probably missing detections of a few low-amplitude pulsators at the faint end, but are otherwise mostly complete.

Fig~\ref{fig:amp_vs_bprp}~(b) shows the same plot as panel (a), except the x-axis is now the Gaia colour \bprp. The corresponding plot in Paper~I (figure 10) showed a curious under-density of stars in the middle of the IS, near $\bprp=0.3$. The maximum amplitude of \dscts\ in this region seemed to have a lower bound, and the maximum white-noise peak of ``not \dsct'' had an upper bound. Here, we note that this region is again relatively sparsely populated compared to either side of the IS, except this feature is now less strong compared to the smaller sample of Paper~I.

We observe in Fig~\ref{fig:amp_vs_bprp}~(b) a rising trend in the $A_1$ of \dscts\ with an increase in \bprp\ (decrease in temperature). This was similarly seen for \dscts\ in the Cep-Her association \citet{Murphy2024}. We also see that $f_1$ is relatively higher for hotter \dscts\ compared to their cooler counterparts, as also seen in \citet{Barcelo-forteza2018}. Physically, this may be due to hotter \dscts\ having thinner partial He II ionization zones, owing to steeper temperature profile. This results in shorter thermal timescales, and hence models predict that driving will preferentially occur for higher-frequency modes \citep{Dupret2004,Xiong2016}. 

\subsection{Integrated power spectral density}\label{section:ipower}

We now introduce a new quantity to describe the amplitude of \dsct\ pulsations, namely the integrated PSD.
Pulsations in \dscts\ are driven by the $\kappa$ mechanism, converting heat energy to kinetic energy and distributing it among modes over a wide range of frequencies with varying amplitudes. This points to the total energy present in pulsations as a more natural quantity to investigate. 

We computed the integrated PSD for each \dsct\ star using the following steps:
\begin{enumerate}
    \item High-pass filter the light curve (divide the original light curve by one that is convolved with a Gaussian kernel of width 1 \cd) to attenuate any leaked power from lower-frequency variability (instrumental artefacts, g-modes, r-modes, rotation, etc.);
    \item Compute the Lomb-Scargle amplitude spectrum and square it to obtain power spectrum (units of ppm$^2$);
    \item Subtract the mean of white noise (in 95--100 \cd\ for 120-s and 200-s cadences and 67--72 \cd\ for 600-s cadence) from the power spectrum;
    \item Multiply by the total length of observation, which is equivalent to dividing by the frequency resolution. This gives the Power Spectral Density ($\text{PSD}=\text{power / unit\ frequency}$, in units of ppm$^2$~/~ \cd);
    \item Integrate PSD over the entire frequency array $\displaystyle\int \text{PSD\ df}$\: to get integrated PSD in ppm$^2$.
\end{enumerate}

We also computed the centroid of the pulsations, referred to as \numax, to encapsulate the characteristic frequency for a \dsct\ star. We used the following steps:
\begin{enumerate}
    \item Mask the Lomb-Scargle amplitude spectrum with the filters  $\left[f \geq 0.66*\fmin\right] \ \&\ \left[A > 5 * \text{median(amplitude)}\right]$;
    \item Square to compute the power spectrum, $P = A^2$;
    \item Compute the power-weighted frequency \citep[see][]{Barcelo-forteza2018, Murphy2024} $\numax = \displaystyle\frac{\sum\limits_i P_i * f_i}{\sum\limits_i P_i}$.
\end{enumerate}
Although we refer to this quantity as ``\numax'' to maintain continuity with the \dsct\ literature \citep[][]{Barcelo-forteza2018, Murphy2024}, it should be noted that it represents the centroid of the power distribution. Only if the power distribution is symmetric will it be equal to the asteroseismic quantity conventionally denoted by \numax\ for solar-like oscillators, which corresponds to the centre of the Gaussian-like power envelope.
Finally, note that the choice of the 5$*$median(amplitude) threshold was determined empirically. Lower thresholds introduced offsets in the measured \numax\ values of low-amplitude \dscts, as the increasing contribution of high-frequency white-noise power began to dominate the calculation.


\section{Results}

\subsection{Integrated PSD}\label{section:integrated_PSD}

\begin{figure*}
\includegraphics[width=\textwidth]{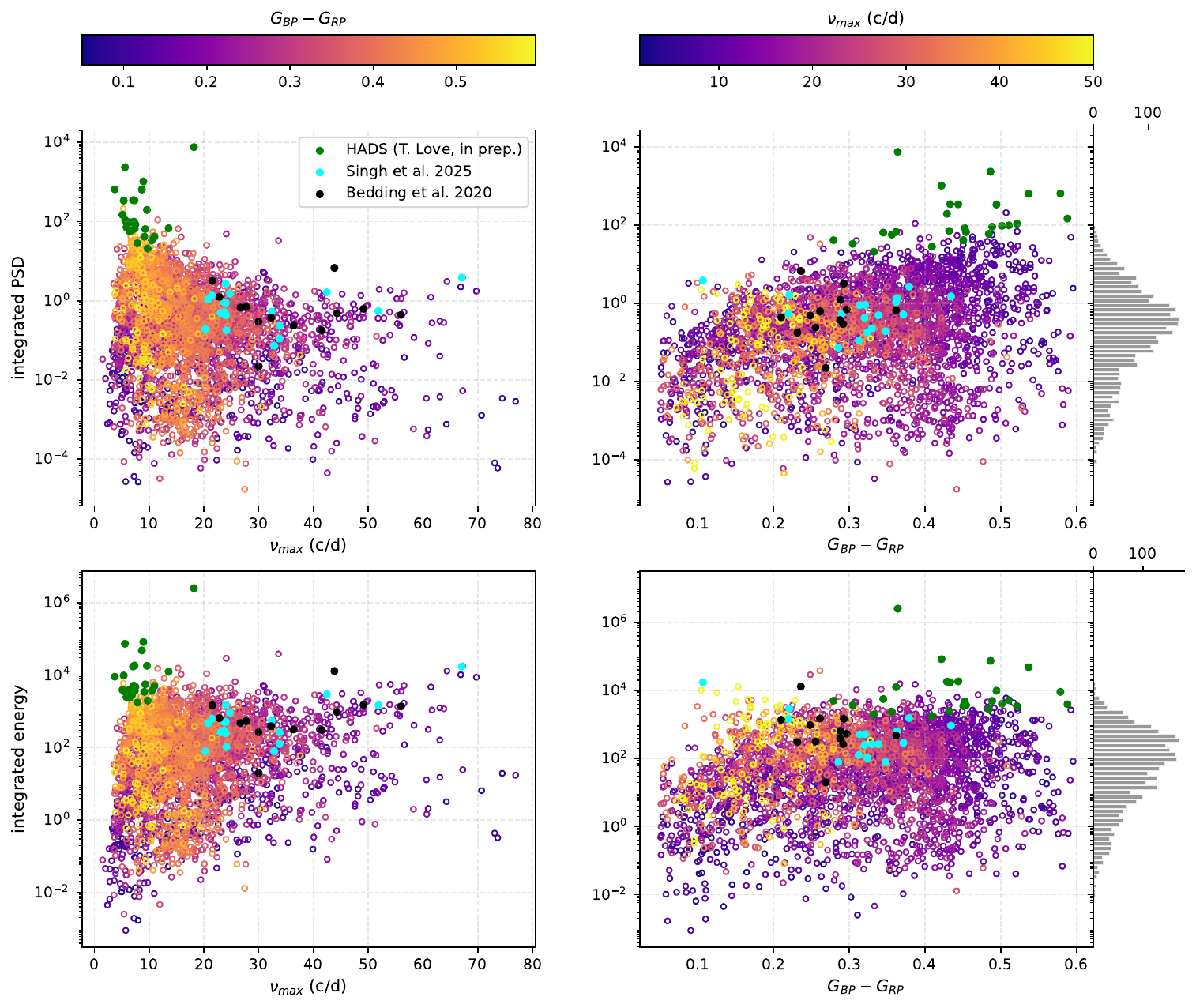}
\caption{(a): integrated PSD vs \numax\ for \dscts. (b): integrated PSD vs \bprp. (c): Integrated energy vs \numax. (d): Integrated energy vs \bprp. Stars in our sample common with T. Love (in prep.) HADS are shown in green, those in common with \citet{Singh2025} are marked with cyan, and those in common with \citet{Bedding2020} are marked with black. Marginalized histograms are shown on the right axes of panels (b) and (d). Points in panels (a) and (c) are colour-coded by \bprp\ and that in panels (b) and (d) are colour-coded by \numax. As very few stars have high \numax\ values we saturate the colourbar at 50 c/d.}
\label{fig:ip_vs_centroid}
\end{figure*}

\begin{figure}
\includegraphics[width=\linewidth]{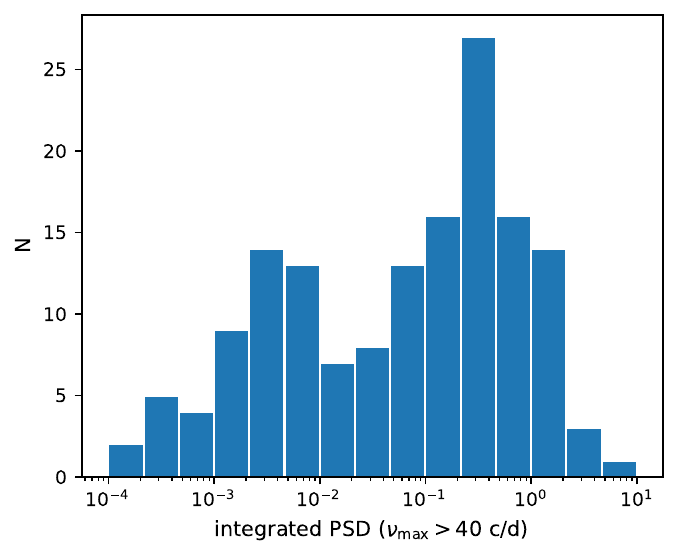}
\caption{Histogram of integrated PSD for \dscts\ whose \numax~$>40$~\cd.}
\label{fig:bimodal}
\end{figure}

\begin{figure}
\includegraphics[width=\linewidth]{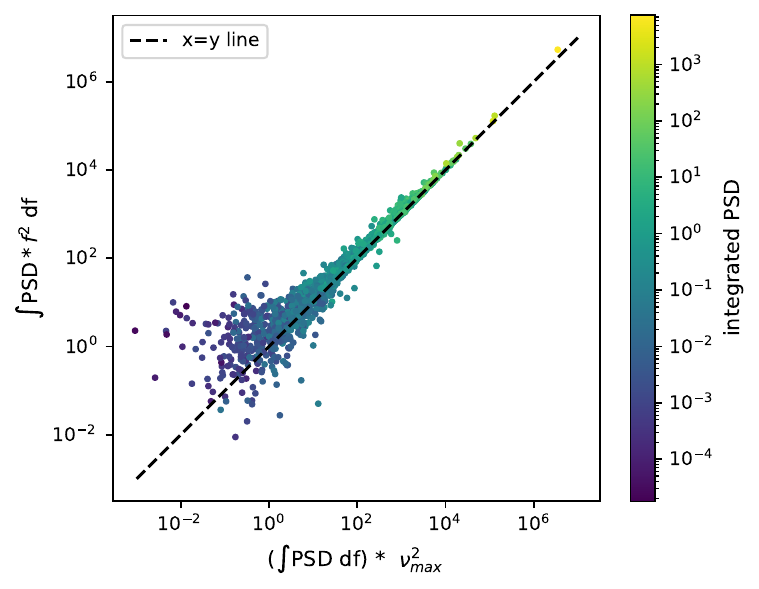}
\caption{Comparison of two different methods of computing energy proxy, see section~\ref{section:integrated_energy}.}
\label{fig:energy_1_vs_2}
\end{figure}

Fig.~\ref{fig:ip_vs_centroid}(a) shows the plot of integrated PSD vs \numax\ for the \dscts\ in our sample. The figure divides the stars into various populations. The population with the highest integrated PSD are the HADS, marked by green circles (see T. Love et al., in prep.). HADS are predominantly non-linear pulsators and show harmonics in their amplitude spectra. For this reason, for the HADS we defined \numax\ to be the frequency of the dominant mode rather than using the centroid of the power. In Fig~\ref{fig:ip_vs_centroid} (a), HADS show a lack of preference for any particular region of the CMD, instead spanning a broad range in colour and magnitude.

Stars to the upper right of Fig.~\ref{fig:ip_vs_centroid}(a), with high integrated PSD and high \numax, are the youngest and least luminous \dscts\ that lie near the ZAMS. Some of these (cyan and black circles) have regular spacing of peaks in the amplitude spectra \citep[][]{Bedding2020, Singh2025}. 

A subpopulation of \dscts\ at high \numax\ occupies the low integrated PSD regime and we found them to be more dispersed in the CMD than their high-power counterparts. We speculate that this behaviour arises out of pulsations being driven, at least in part, by turbulent pressure \citep[][]{Antoci2014, Antoci++2019} rather than the classical $\kappa$ mechanism. 
As the driving layer is shallow, closer to the surface, and the mechanism is stochastic rather than coherent, it naturally shifts power toward higher frequencies (\numax) while suppressing the total integrated PSD. We identify an under-density separating the two high-\numax\ populations, which is shown more clearly in Fig.~\ref{fig:bimodal}. Given the completeness of our sample, this feature is unlikely to be a statistical artefact and may reflect a transition between dominant driving regimes.

Fig.~\ref{fig:ip_vs_centroid}~(b) shows integrated PSD plotted against \bprp, illustrating the rising summed power as we traverse across the IS from hotter to cooler edge. Although this plot is similar to Fig.~\ref{fig:amp_vs_bprp}~(b), the trend of rising power is much more evident when integrated PSD is considered instead of merely the dominant mode, $A_1$. Colour-coding by \numax\ shows that high-frequency pulsators are preferentially hotter, just as in Fig.~\ref{fig:amp_vs_bprp}~(b). This is consistent with previous observations \citep{Balona+Dziembowski2011, Barcelo-forteza2018, Ziaali2019} and with theoretical expectations of higher overtones being excited in hotter stars \citep{Dziembowski1997, Houdek1999, Xiong2001, Dupret2004, Xiong2016}.

The majority of our \dsct\ sample pulsates in the $\numax$ range of $\simeq 15-20\ \cd$ and at intermediate power. This population of \dscts\ also does not appear to cluster in any specific region of either the (IS) or the (CMD). The distribution of stars is seemingly bounded by a sloping envelope extending from low \numax\ and high power to high \numax\ and intermediate power. This leads us to ask -- is there a trade-off where modes compete for the driving energy reservoir with either high amplitudes or high frequencies, but not both?

\subsection{Integrated energy}\label{section:integrated_energy}
In Fig.~\ref{fig:ip_vs_centroid}~(c) and (d), we remake panels (a) and (b) by modifying the y-axis to show an estimate of the net pulsation energy \citep[see also][]
{Uytterhoeven2011}. For a simple harmonic oscillator, $E \propto \omega^2A^2$, where $\omega$ is the oscillation frequency and $A$ the amplitude. To convert integrated PSD to energy, we therefore simply multiplied by $\numax^2$. A stricter definition of energy would include the frequency term within the integral, i.e., $\int\text{PSD}\ f^2\ \text{df}$, so that the energy of each mode is calculated using its own frequency. However, Fig~\ref{fig:energy_1_vs_2} shows that it tends to overestimate pulsation energy in low-amplitude \dscts.  This is because the $f^2$ term magnifies the high-frequency region, leading to extra contribution from white-noise into the calculation if the white noise has not been subtracted precisely. Our definition sidesteps this problem, but it does assume that all the pulsation energy is concentrated near \numax. In practice, for intermediate- and high-power \dscts, the results from these two methods agree (from around $\sim10^1$ in Fig.~\ref{fig:energy_1_vs_2}).

Fig.~\ref{fig:ip_vs_centroid}~(c) and (d) show that after converting integrated PSD to energy, the rising trend fades and the observed pulsation energy broadly flattens across the population. This suggests an upper limit to the efficiency with which the $\kappa$ mechanism converts thermal energy into pulsation energy. In panel (d), toward the blue edge of the IS, the inferred energy decreases, consistent with reduced driving efficiency as the convective envelope thins and the He~II ionization zone shifts outward. HADS depart from this behaviour. As large-amplitude, non-linear pulsators, their pulsations are not well described by a linear oscillator model. 

As a caveat we note that this proxy for energy we derived is based on a star's brightness variations, which are approximately proportional to the physical displacement of the star's surface \citep[e.g.,][]{Kjeldsen1995}. To quantify relative pulsation strengths in \dscts, we calculate the energy proxy based on observed power. \citep[see][]{Uytterhoeven2011}.

\subsection{Colour-Magnitude diagram}\label{section:CMD}

\begin{figure*}
\includegraphics[width=\linewidth]{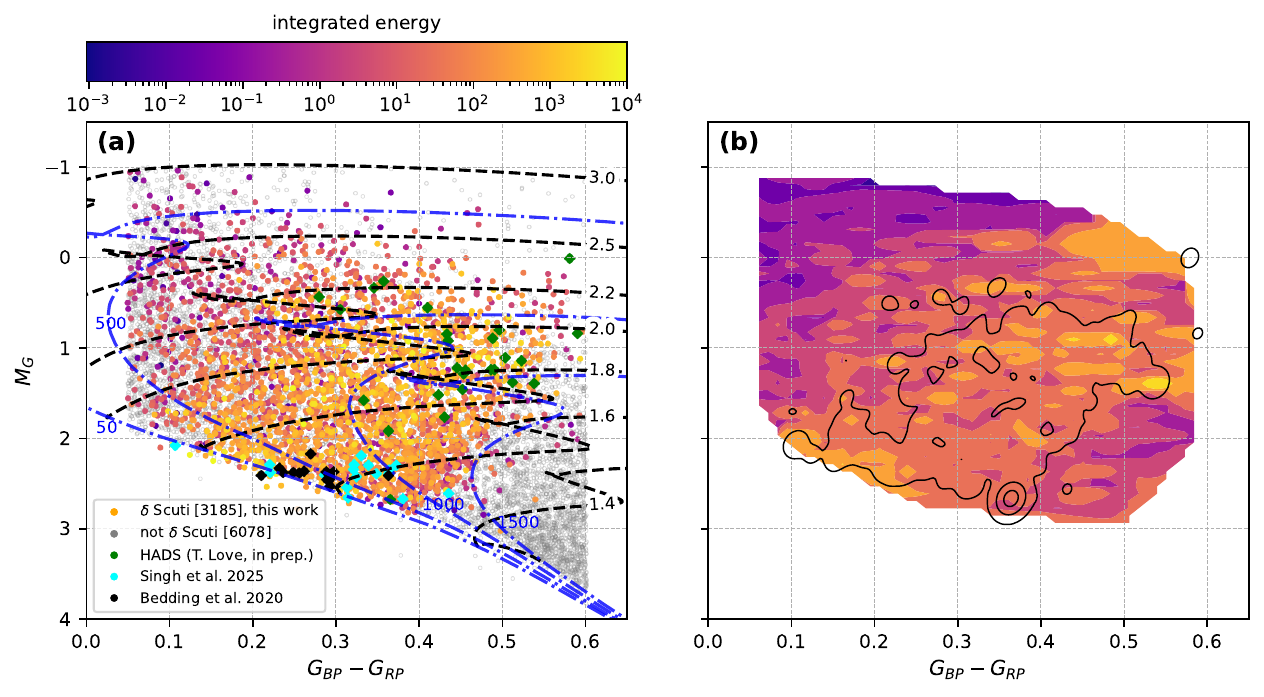}
\caption{(a): Colour-magnitude diagram of our sample, with \dscts\ colour-coded by integrated pulsation energy (see Sec.~\ref{section:integrated_energy}). The curves show MIST evolutionary tracks (blacked dashed lines, labelled in solar mass units) and isochrones (blue dot-dashed lines, labelled in Gyr), both for a rotation rate of $\nu/\nu_{\text{crit}} = 0.4$. HADS are shown in green, those in common with \citet{Singh2025} are marked with cyan, and those in common with \citet{Bedding2020} are marked with black. (b): 2D-contour plot of the \dscts. We interpolated the distribution of \dscts\ in (a) according to their integrated energy values, on the CMD grid. The contour lines are energy-weighted kernel density estimate (KDE). That is, contour lines show regions where $\left( \text{integrated energy} \cdot \text{number of}\ \delta \ \text{Scuti stars} \right)$ peaks.
}
\label{fig:CMD}
\end{figure*}

Fig.~\ref{fig:CMD}~(a) shows the colour-magnitude diagram (CMD), with \dscts\ colour-coded by their integrated energy (see section~\ref{section:integrated_energy}), and all other stars are shown in grey. Evolutionary tracks are shown as black dashed lines, and isochrones as blue dot-dashed lines, both obtained from MIST~\footnote{\url{https://mist.science}} \citep{Choi2016}. The red edge of the instability strip appears to have a sharp boundary, whereas the diffuse blue edge shows a sparser \dsct\ population compared to the bulk of the instability strip. 

\begin{figure}
\includegraphics[width=\linewidth]{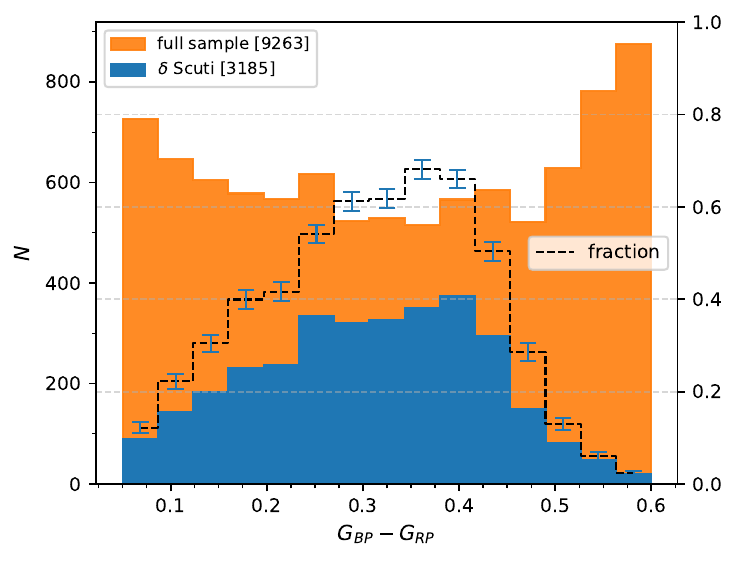}
\caption{Pulsation fraction as a function of \bprp.  The left axis is the number of stars in the histograms (blue for \dscts\ and orange for the full sample), while the right axis denotes the pulsation fraction (black dashed histogram, with error bars). }
\label{fig:pulsfrac_bprp}
\end{figure}

The CMD clearly illustrates the longstanding question: Why do only some stars in the IS pulsate as \dscts\ \citep[see][]{murphyetal2015a, Murphy2019, Gootkin2024, Mani2025, Berry2025}? 
Fig.~\ref{fig:pulsfrac_bprp} shows that even in the middle of the IS, near $\bprp=0.3$, the peak pulsator fraction is merely 70\%. The pulsator fraction reduces slowly towards the hot edge, but sharply drops towards the cooler edge. This drop was also noted in the samples studied by \citet[][]{Murphy2019, Mani2025}. \citet{Dupret2005} showed using time-dependent convection models how convection at the cooler edge of the IS acts to efficiently damp pulsations in cool stars. 

In Fig.~\ref{fig:CMD}~(b), for an overall perspective, we smooth and interpolate the distribution of \dscts\ in (a). Warmer regions denote populations with higher integrated energy values. Pulsation energy declines on either side of the (IS) as denoted by the purple regions. Kernel density estimate (KDE) lines weighted by integrated energy is overplotted, indicating regions of high concentration of large pulsation energy \dscts. One such region is populated by young \dscts\ near the ZAMS (Zero Age Main Sequence) marked by cyan and black points in (a). These are the high frequency \dscts\ with regularly spaced peaks in their Fourier amplitude spectra. Whereas near high luminosity region, KDE identifies the HADS population, marked by green points in panel (a). Integrated energy for our \dsct\ sample is discussed in more detail in section~\ref{section:integrated_energy}.

\subsection{Period--Luminosity relation}\label{section:PL} 

Many previous papers studied the Period--Luminosity (P--L) relation for \dscts\ \citep[e.g.,][]{McNamara2011, Ziaali2019, Barac++2022, Martinez-Vazquez++2022, Gaia-De-Ridder++2023, Soszynski++2023, PLreview2024, Vasigh++2024, Guo-Fangzhou++2025, Jia++2025}. Using the P--L relation, \dscts\ can potentially be used as tracers of the disk structure of the Milky Way \citep[with young population, see][]{Wang++2015} or distances to globular clusters \citep[with older population, for e.g.,][]{McNamara2011}. More importantly for this work, we can use the P--L relation to predict the expected p-mode frequencies for a given star and hence decide whether it is a \dsct\ pulsator (see Sec.~\ref{section:data analysis}).

\begin{figure*}
\includegraphics[width=\linewidth]{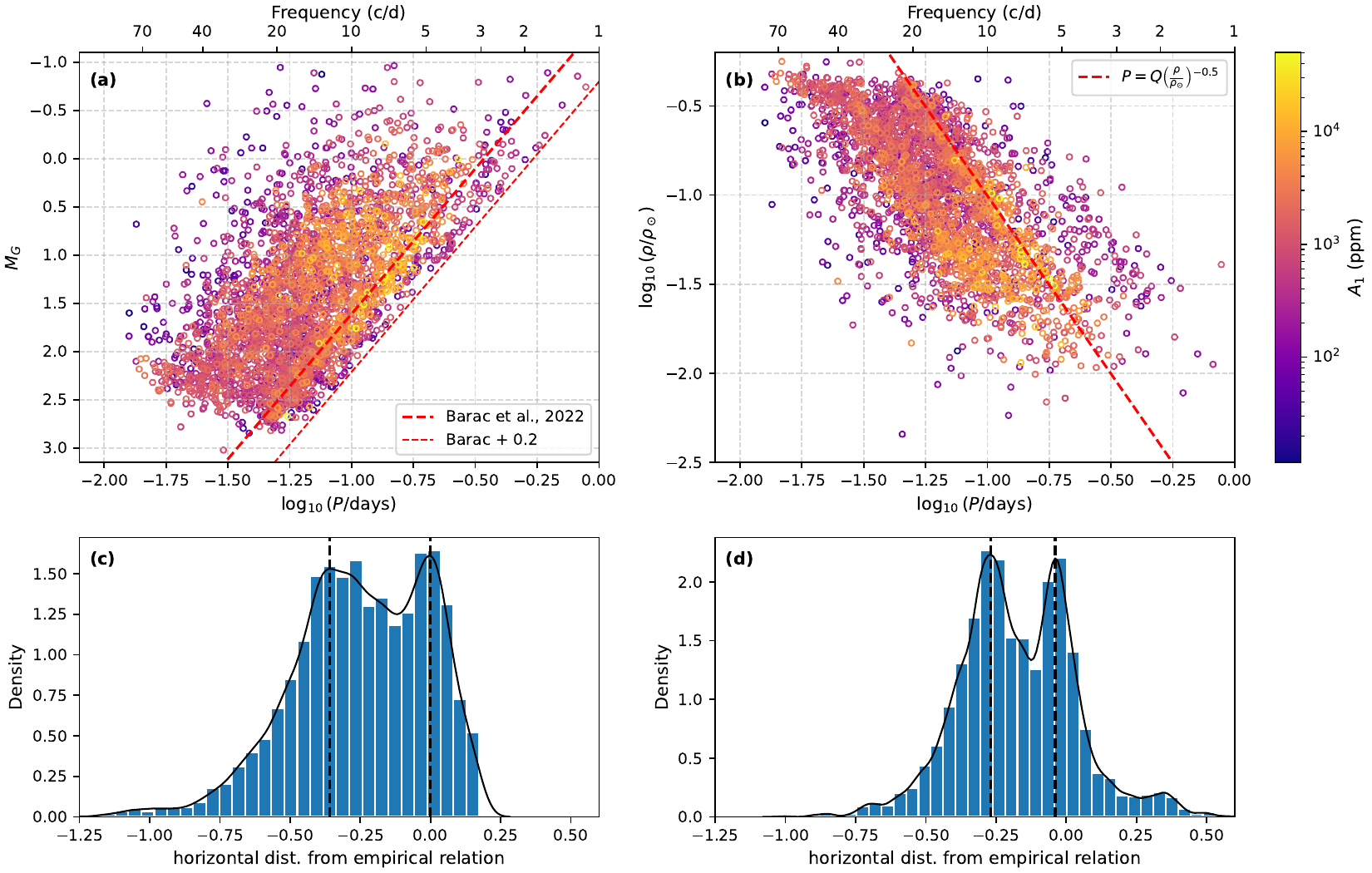}
\caption{(a): Period--Luminosity relation for \dscts. Red dashed line denotes the fundamental mode \fmin\ \citep[][: $M_G = -3.01\log_{10}P -1.4 $]{Barac++2022}. Red dotted line denotes \fmin\ shifted to the right by 0.2 in ${\rm{log}_{10}} (P/\rm{days})$. (b): Period--density relation for \dscts. Red dashed line denotes the scaling relation for fundamental period vs density \citep[][: $P = Q \left( \frac{\rho}{\rho_\odot} \right)^{-0.5}$]{Barac++2022}. Stars in (a) and (b) are colour-coded by $A_1$, and the colourbar is saturated at $\mathrm{5e4}$. (c): kernel density estimate (KDE) overlaid on histogram of horizontal distances of stars from \fmin\ line shown in (a). (d): kernel density estimate (KDE) overlaid on histogram of horizontal distances of stars from the fundamental mode scaling relation line shown in (b).}
\label{fig:PL}
\end{figure*}

We plot the P--L relation for our \dscts\ in Fig.~\ref{fig:PL}~(a), based on the dominant peak in each spectrum (Sec.~\ref{section: A_1 and f_1}). The red dashed line shows the empirical relation in \citet{Barac++2022}: $M_G = -3.01 \log_{10}(P/\text{d}) -1.4 $, where $P$ is the period. We see an over-density of \dscts\ close to this line, which corresponds to the radial fundamental mode (frequency~\fmin). Fig.~\ref{fig:PL}~(c) shows the histogram of horizontal distances of \dscts\ from \fmin. The black dashed line at around $-0.3\simeq\log_{10}(1/2)$  is near half the fundamental mode period, as found by \citet{Ziaali2019, Barac++2022}. This indicates that many \dscts\ might also pulsate in third or fourth overtone pulsations \citep[][]{Jayasinghe2020, Soszynski++2023}.

As mentioned in section~\ref{section:data analysis}, we found several \dscts\ to pulsate slightly below their \fmin. We can attribute this to various factors, such as an intrinsic spread in the relation, incorrect parallax measurements, mode splitting in very fast rotating \dscts, presence of mixed-modes \citep[][]{christensen_dalsgaard_2000}, and a hybrid \dsct--\gdor\ star \citep[][]{Grigahcene2010, Antoci++2019, Guzik2021}. Thus we classified as \dscts\ those whose spectrum had at least one strong peak (S/N $>$ 5) above 0.66\fmin. This threshold is marked by the red dotted line in Fig~\ref{fig:PL}~(a), a distance of $\log_{10}P = 0.2$ from \fmin.

\subsection{Period--density relation}

To a good approximation, the frequency of a given p mode (such as the radial fundamental) scales with the mean stellar density:
$\nu \propto \sqrt{\rho}$. We can write this as $$P=Q\left(\frac{\rho}{\rho_{\odot}}\right)^{-0.5},$$ where $P$ is the mode period and $Q$ is the so-called pulsation constant \citep[e.g.,][]{Catelan+Smith2015}. Fig.~\ref{fig:PL}~(b) shows the distribution of dominant period of the \dscts\ in our sample vs their densities. The densities were estimated using the software \texttt{isoclassify} \citep[see][]{Huber2017}, which outputs a posterior distribution of stellar parameters given a set of observables (in our case, we used \bprp, $G$ magnitude, parallax, and the position co-ordinates RA and DEC for each \dsct\ star). In the figure, we observe two dominant ridges in the distribution, corresponding to fundamental mode shown by the red-dashed line with $Q=0.0315$d, and the second ridge at a horizontal distance of 0.3 (in $\log$ scale) away from the fundamental period. This second ridge corresponds to the one observed in panel (a), with mode period being half that of the fundamental mode, but is seen even more clearly (see histogram in Fig.~\ref{fig:PL}~d). A small fraction of points lie well to the right of the fundamental mode relation. We have confirmed that these are genuine \dscts\ but that there is a strong peak just above the 0.66\fmin\ threshold, which is plotted here. These peaks are probably from low-frequency variability and are not p~modes.

\section{Conclusion}

Our goal in this paper was to propose a simple and effective calculation of mode power and total mode energy, based on previous efforts in solar-like oscillators. This comprehensive approach provides a more physically meaningful description of pulsations in multi-periodic stars than characterizing them solely by their dominant amplitude and corresponding frequency. Fig.~\ref{fig:ip_vs_centroid} (a) shows the presence of a structure in the distribution of \dsct\ mode power and \numax. There may be two subpopulations of high \numax\ \dscts, driven dominantly by $\kappa$-mechanism and turbulent pressure. Also, our calculation of mode energy in panels (c) and (d) showed a plateau in the upper part of the figure, suggesting a limit to the driving efficiency in these stars.

Future work will be dedicated to the analysis of mode power and energy in other multiperiodic pulsators such as
$\gamma$~Doradus, SPB (Slowly Pulsating B-type), and $\beta$~Cephei stars. Especially in $\gamma$~Doradus stars, our understanding of driving efficiency can be greatly improved by independently analyzing pulsation energy in different mode families (such as prograde / retrograde sectoral / tesseral modes, purely inertial / gravito-inertial modes, etc.)   

\section{Acknowledgements} We thank the referee for their useful comments. We gratefully acknowledge support from the Australian Research Council through Future Fellowship FT210100485, and Laureate Fellowship FL220100117.
This work has made use of data from the European Space Agency (ESA) mission {\em Gaia}, (\url{https://www.cosmos.esa.int/gaia}), 
processed by the {\em Gaia} Data Processing and Analysis Consortium (DPAC, \url{https://www.cosmos.esa.int/web/gaia/dpac/consortium}). 
We are grateful to the entire Gaia and \tess\ teams for providing the data used in this paper.
This work made use of several publicly available {\tt python} packages: {\tt astropy} \citep{astropy:2013,astropy:2018}, 
{\tt lightkurve} \citep{lightkurve2018},
{\tt matplotlib} \citep{matplotlib2007}, 
{\tt numpy} \citep{numpy2020}, and 
{\tt scipy} \citep{scipy2020}.

\section*{Data Availability}

The \tess\ data underlying this article are available at the MAST Portal (Barbara A. Mikulski Archive for Space Telescopes), at \url{https://mast.stsci.edu/portal/Mashup/Clients/Mast/Portal.html}

\ifarxiv
    \input{output.bbl} 
\else
    \bibliographystyle{mnras}
    \bibliography{references}
\fi

\label{lastpage}

\end{document}